\documentclass[10pt]{iopart}

\usepackage{graphicx}
\usepackage{dcolumn}
\usepackage{bm}
\usepackage{color}

\begin{document}

\title[Optimization of DIII-D reverse shear configurations]{Subdominant modes and optimization trends of DIII-D reverse magnetic shear configurations}


\author{J. Varela}
\ead{rodriguezjv@ornl.gov}
\address{Oak Ridge National Laboratory, Oak Ridge, Tennessee 37831-8071, USA}
\author{D. A. Spong}
\address{Oak Ridge National Laboratory, Oak Ridge, Tennessee 37831-8071, USA}
\author{M. Murakami}
\address{Oak Ridge National Laboratory, Oak Ridge, Tennessee 37831-6069, USA}
\author{L. Garcia}
\address{Universidad Carlos III de Madrid, 28911 Leganes, Madrid, Spain}
\author{E. D'Azevedo}
\address{Oak Ridge National Laboratory, Oak Ridge, Tennessee 37831-6069, USA}
\author{M.A. Van Zeeland}
\address{General Atomics, PO Box 85608 San Diego, California 92186-5608, USA}
\author{S. Munaretto}
\address{General Atomics, San Diego, CA, 92186, USA}

\date{\today}

\begin{abstract}
Alfv\' en Eigenmodes (AE) and magneto-hydrodynamic (MHD) modes are destabilized in DIII-D reverse magnetic shear configurations and may limit the performance of the device. We use the reduced MHD equations in a full 3D system, coupled with equations of density and parallel velocity moments for the energetic particles (with gyro-fluid closures) as well as the geodesic acoustic wave dynamics, to study the properties of instabilities observed in DIII-D reverse magnetic shear discharges. The aim of the study consists in finding ways to avoid or minimize MHD and AE activity for different magnetic field configurations and neutral beam injection (NBI) operational regimes. The simulations show at the beginning of the discharge, before the reverse shear region is formed, a plasma that is AE unstable and marginally MHD stable. As soon as the reverse shear region appears, ideal MHD modes are destabilized with a larger growth rate than the AEs. Both MHD modes and AEs coexist during the discharge, although the MHD modes are more unstable as the reverse shear region deepens. The simulations indicate the destabilization of Beta induced AE (BAE), Toroidal AE (TAE), Elliptical AE (EAE) and Reverse Shear AE (RSAE) at different phases of the discharges, showing a reasonable agreement between the frequency range of the dominant modes in the simulations and the diagnostic measurements. A further analysis of the NBI operational regime indicates that the AE stability can be improved if the NBI injection is off axis, because on-axis injection leads to AEs with larger growth rate and frequency. In addition, decreasing the beam energy or increasing the NBI relative density (the ratio between the energetic particle and thermal plasma density) leads to AEs with larger growth rate and frequency, so an NBI operation in the weakly resonant regime requires higher beam energies than in the experiment ($V_{th,f} / V_{A0} > 0.3$). The MHD linear stability can be also improved if the reverse shear region and the q profile near the magnetic axis are in between the rational surfaces $q=2$ and $q=1$, particularly if there is a region in the core with negative shear, avoiding a flat q profile near the magnetic axis. The simulations also show a smooth transition between MHD modes and low frequency AE, no critical $\beta_{f}$, pointing out an overlap between MHD and AE activity for modes with frequencies lower than $30$ kHz. This is in the range of Beta Acoustic Alfven Eigenmodes (BAAE) and BAE.
\end{abstract}

%
%
%
%
%


\vspace{2pc}

This manuscript has been authored by UT-Battelle, LLC under Contract No. DE-AC05- 00OR22725 with the U.S. Department of Energy. The United States Government retains and the publisher, by accepting the article for publication, acknowledges that the United States Government retains a non-exclusive, paid-up, irrevocable, world-wide license to publish or reproduce the published form of this manuscript, or allow others to do so, for United States Government purposes. The Department of Energy will provide public access to these results of federally sponsored research in accordance with the DOE Public Access Plan (http://energy.gov/downloads/doe-public-access-plan).

\maketitle

\ioptwocol

\section{Introduction \label{sec:introduction}}

Reverse magnetic shear discharges are a possible operational scenario of nuclear fusion reactors because these configurations have good MHD stability properties and energy confinement at high $\beta$ \cite{1,2,3,4,5,6}, although Alfv\' en Eigenmode stability and energetic particle confinement are still an open issue  \cite{7,8,9,10,11,12,13}. MHD stability of interchange and ballooning modes is improved in high $\beta$ discharges by combining reverse shear operation (the plasma is in the second stable regime \cite{14,15}) and fast toroidal rotation \cite{16}. In addition, the energy confinement is improved in reverse shear configurations due to the stabilization of drift-type micro-instabilities \cite{17}. Also, significant bootstrap current can be generated, lowering the current drive requirement for steady-state operation \cite{18,19}. 

Reverse magnetic shear discharges are extensively studied in the DIII-D device \cite{20,21,22} showing a large fraction of bootstrap current \cite{23,24,25} and improved MHD stability \cite{26,27,28}. These configurations have been selected as a base line scenario for ITER and DEMO \cite{29,30,31}, although an enhancement of the energetic particle transport by unstable Alfven Eigenmodes was measured \cite{32,33,34}.

The transport of energetic neutral beam ions, particles heated using ion cyclotron resonance heating (ICRF) and fusion produced alpha particles can be  enhanced by energetic particles driven instabilities \cite{35,36,37}. Experiments in TFTR, JET and DIII-D tokamaks as well as LHD and W7-AS stellarators identified a decrease in the device performance caused by the destabilization of Alfven Eigenmodes \cite{38,39,40,41,42,43}, leading to an increase of particle and diffusive loses if the transit, bounce or drift frequency of the energetic particles resonate with AE mode frequencies. 

Super-Alfv\' enic alpha particles and energetic particles destabilize Alfv\' en Eigenmodes (AE) in the spectral gaps of the shear Alfv\' en continua \cite{44,45}; these have been seen in several DIII-D configurations \cite{46,47,48,49}. Periodic variations of the Alfv\' en speed destabilize different AE families, for example toroidicity induced Alfv\' en Eigenmodes (TAE) coupling $m$ with $m+1$ modes (m is the poloidal mode) \cite{50,51}, ellipticity induced Alfv\' en Eigenmodes (EAE) coupling $m$ with $m+2$ modes \cite{52,53} or noncircularity induced Alfv\' en Eigenmodes (NAE) coupling $m$ with $m+3$ or higher \cite{54,55}. Other example of AE are the beta induced Alfv\' en Eigenmodes (BAE) driven by compressibility effects \cite{56}, Reversed-shear Alfv\' en Eigenmodes (RSAE) due to local maxima/minima in the safety factor $q$ profile \cite{32,57} and Global Alfv\' en Eigenmodes (GAE) observed in the minimum of the Alfv\' en continua \cite{58,59}. Furthermore, current driven instabilities as internal kinks \cite{60,61} or pressure gradient driven instabilities as interchange and ballooning modes \cite{62} can be kinetically destabilized. textcolor{red}{Please, see the Appendix for further details about the instability characteristics and eigenfunction structure, as well as a comparison with the electron temperature fluctuations (ECE) data measured during the discharges.}

DIII-D has eight neutral beam injectors (NBI), six sources injected in the midplane (on axis) and 2 injected downwards at an angle (off axis) to heat, fuel and drive current in the plasma. Six sources are co-injected (direction of the plasma current) with two tilted sources and 2 source are counter-injected (opposite to the plasma current). The NBI injects deuterium with a beam energy of $80$ keV ($2.25$ MW source) in a deuterium plasma. Strong NBI heating leads to the destabilization of AE as  TAE \cite{63}, EAE \cite{64}, NAE \cite{65}, BAE \cite{66}, RSAE \cite{67} and GAE \cite{68}. DIII-D experiments measured a larger transport and enhanced energetic particle losses linked to unstable AE \cite{69,70,71}.

The aim of the present study is to identify optimization pathways to improve MHD and AE linear stability of DIII-D reverse magnetic shear discharges, comparing simulation results and experimental observations. We analyze the effect of the magnetic field configuration and NBI operation regime (beam energy, injection intensity and deposition profile) on MHD and AE stability. We identify dominant and sub-dominant MHD and AE modes at different phases in a selection of three reverse magnetic shear DIII-D discharges. In addition, we study the overlapping between low frequency AE (frequency smaller than $40$ kHz) and MHD modes, as well as the effect of the energetic particle forcing on the MHD modes growth rate and frequency.

This analysis is performed using the FAR3D code \cite{72,73,74}, with extensions to include the moment equations of the energetic ion density and parallel velocity \cite{75,76} allowing treatment of linear wave-ion resonances. The numerical code solves the reduced non-linear resistive MHD equations adding the Landau damping/growth (wave-particle resonance effects) and geodesic acoustic waves (parallel momentum response of the thermal plasma) \cite{57}. The Landau closure relations (Landau-closure models can be calibrated by more complete kinetic models \cite{57}). The simulations are based on an equilibria calculated by the VMEC code \cite{77}.

This paper is organized as follows. The model equations, numerical scheme and equilibrium properties are described in section \ref{sec:model}. The study of dominant and subdominant modes at different phases of reverse magnetic shear discharges is shown in section \ref{sec:main}. The optimization trends to minimize AE and MHD activity modifying the NBI operational regime and the magnetic field configuration are analyzed in section \ref{sec:Opt}. Studies of the interaction between MHD modes and low frequency AE is presented in section \ref{sec:MHD/AE}. Finally, the conclusions of this paper are presented in section \ref{sec:conclusions}.

\section{Equations and numerical scheme \label{sec:model}}

From the full MHD equations we can derive a reduced set of three equations that follow the evolution of the pressure $p$, the stream function proportional to the electrostatic potential $\Phi$ and the perturbation of the poloidal flux $\psi$ while retaining the toroidal angle variation $\zeta$ (exact three-dimensional equilibrium) \cite{78}. The equations for the background or thermal plasma evolution are complemented by moments of the energetic ions to add the effect of the wave-particle interaction \cite{79}, namely the energetic particle density ($n_{f}$) and velocity moments parallel to the magnetic field lines ($v_{||f}$). The coefficients of the closure relation are selected to match a two-pole approximation of the plasma dispersion function.   

The reduced equations are based on the following assumptions: high aspect ratio, medium $\beta$ (of the order of the inverse aspect ratio $\varepsilon=a/R_0$), small variation of the fields and small resistivity. The magnetic field and plasma velocity perturbations are defined as:
\begin{equation}
 \mathbf{v} = \sqrt{g} R_0 \nabla \zeta \times \nabla \Phi, \quad\quad\quad  \mathbf{B} = R_0 \nabla \zeta \times \nabla \psi,
\end{equation}

The equations, in dimensionless form, are
\begin{equation}
\frac{\partial \tilde \psi}{\partial t} =  \sqrt{g} B \nabla_\| \Phi  + \eta \varepsilon^2 J \tilde J^\zeta
\end{equation}
\begin{eqnarray} 
\frac{{\partial \tilde U}}{{\partial t}} =  -\epsilon v_{\zeta,eq} \frac{\partial U}{\partial \zeta} \nonumber\\
+ S^2 \left[{ \sqrt{g} B \nabla_\| J^\zeta - \frac{\beta_0}{2\varepsilon^2} \sqrt{g} \left( \nabla \sqrt{g} \times \nabla \tilde p \right)^\zeta }\right]   \nonumber\\
-  S^2 \left[{\frac{\beta_f}{2\varepsilon^2} \sqrt{g} \left( \nabla \sqrt{g} \times \nabla \tilde n_f \right)^\zeta }\right] 
\end{eqnarray} 
\begin{eqnarray}
\label{pressure}
\frac{\partial \tilde p}{\partial t} = -\epsilon v_{\zeta,eq} \frac{\partial p}{\partial \zeta} + \frac{dp_{eq}}{d\rho}\frac{1}{\rho}\frac{\partial \tilde \Phi}{\partial \theta} \nonumber\\
 +  \Gamma p_{eq}  \left[{ \sqrt{g} \left( \nabla \sqrt{g} \times \nabla \tilde \Phi \right)^\zeta - \nabla_\|  v_{\| th} }\right] 
\end{eqnarray} 
\begin{eqnarray}
\label{velthermal}
\frac{{\partial \tilde v_{\| th}}}{{\partial t}} = -\epsilon v_{\zeta,eq} \frac{\partial v_{||th}}{\partial \zeta} -  \frac{S^2 \beta_0}{n_{0,th}} \nabla_\| p 
\end{eqnarray}
\begin{eqnarray}
\label{nfast}
\frac{{\partial \tilde n_f}}{{\partial t}} = -\epsilon v_{\zeta,eq} \frac{\partial n_{f}}{\partial \zeta} - \frac{S  v_{th,f}^2}{\omega_{cy}}\ \Omega_d (\tilde n_f) - S  n_{f0} \nabla_\| v_{\| f}   \nonumber\\
- \varepsilon^2  n_{f0} \, \Omega_d (\tilde \Phi) + \varepsilon^2 n_{f0} \, \Omega_* (\tilde  \Phi) 
\end{eqnarray}
\begin{eqnarray}
\label{vfast}
\frac{{\partial \tilde v_{\| f}}}{{\partial t}} = -\epsilon v_{\zeta,eq} \frac{\partial v_{||f}}{\partial \zeta}  -  \frac{S  v_{th,f}^2}{\omega_{cy}} \, \Omega_d (\tilde v_{\| f}) \nonumber\\
- \left( \frac{\pi}{2} \right)^{1/2} S  v_{th,f} \left| \nabla_\|  v_{\| f}  \right| \nonumber\\
- \frac{S  v_{th,f}^2}{n_{f0}} \nabla_\| n_f + S \varepsilon^2  v_{th,f}^2 \, \Omega_* (\tilde \psi) 
\end{eqnarray}
Here, $U =  \sqrt g \left[{ \nabla  \times \left( {\rho _m \sqrt g {\bf{v}}} \right) }\right]^\zeta$ is the vorticity and $\rho_m$ the ion and electron mass density. The toroidal current density $J^{\zeta}$ is defined as:
\begin{eqnarray}
J^{\zeta} =  \frac{1}{\rho}\frac{\partial}{\partial \rho} \left(-\frac{g_{\rho\theta}}{\sqrt{g}}\frac{\partial \psi}{\partial \theta} + \rho \frac{g_{\theta\theta}}{\sqrt{g}}\frac{\partial \psi}{\partial \rho} \right) \nonumber\\
- \frac{1}{\rho} \frac{\partial}{\partial \theta} \left( \frac{g_{\rho\rho}}{\sqrt{g}}\frac{1}{\rho}\frac{\partial \psi}{\partial \theta} + \rho \frac{g_{\rho \theta}}{\sqrt{g}}\frac{\partial \psi}{\partial \rho} \right)
\end{eqnarray}
 $v_{\zeta,eq}$ is the equilibrium toroidal rotation and $v_{||th}$ is the parallel velocity of the thermal particles. $n_{f}$ is normalized to the density at the magnetic axis $n_{f_{0}}$, $\Phi$ to $a^2B_{0}/\tau_{R}$ and $\Psi$ to $a^2B_{0}$. All lengths are normalized to a generalized minor radius $a$; the resistivity to $\eta_0$ (its value at the magnetic axis); the time to the resistive time $\tau_R = a^2 \mu_0 / \eta_0$; the magnetic field to $B_0$ (the averaged value at the magnetic axis); and the pressure to its equilibrium value at the magnetic axis. The Lundquist number $S$ is the ratio of the resistive time to the Alfv\' en time $\tau_{A0} = R_0 (\mu_0 \rho_m)^{1/2} / B_0$. $\rlap{-} \iota$ is the rotational transform, $v_{th,f} = \sqrt{T_{f}/m_{f}}$ the energetic particle thermal velocity normalized to the Alfv\' en velocity in the magnetic axis $v_{A0}$ and $\omega_{cy}$ the energetic particle cyclotron frequency times $\tau_{A0}$. $q_{f}$ is the charge, $T_{f}$ the temperature and $m_{f}$ the mass of the energetic particles. The $\Omega$ operators are defined as:
\begin{eqnarray}
\label{eq:omedrift}
\Omega_d = \frac{1}{2 B^4 \sqrt{g}}  \left[  \left( \frac{I}{\rho} \frac{\partial B^2}{\partial \zeta} - J \frac{1}{\rho} \frac{\partial B^2}{\partial \theta} \right) \frac{\partial}{\partial \rho}\right] \nonumber\\
-   \frac{1}{2 B^4 \sqrt{g}} \left[ \left( \rho \beta_* \frac{\partial B^2}{\partial \zeta} - J \frac{\partial B^2}{\partial \rho} \right) \frac{1}{\rho} \frac{\partial}{\partial \theta} \right] \nonumber\\ 
+ \frac{1}{2 B^4 \sqrt{g}} \left[ \left( \rho \beta_* \frac{1}{\rho} \frac{\partial B^2}{\partial \theta} -  \frac{I}{\rho} \frac{\partial B^2}{\partial \rho} \right) \frac{\partial}{\partial \zeta} \right]
\end{eqnarray}

\begin{eqnarray}
\label{eq:omestar}
\Omega_* = \frac{1}{B^2 \sqrt{g}} \frac{1}{n_{f0}} \frac{d n_{f0}}{d \rho} \left( \frac{I}{\rho} \frac{\partial}{\partial \zeta} - J \frac{1}{\rho} \frac{\partial}{\partial \theta} \right) 
\end{eqnarray}
Here the $\Omega_{d}$ operator is constructed to model the average drift velocity of a passing particle and $\Omega_{*}$ models its diamagnetic drift frequency. We also define the parallel gradient and curvature operators:
\begin{equation}
\label{eq:gradpar}
\nabla_\| f = \frac{1}{B \sqrt{g}} \left( \frac{\partial \tilde f}{\partial \zeta} +  \rlap{-} \iota \frac{\partial \tilde f}{\partial \theta} - \frac{\partial f_{eq}}{\partial \rho}  \frac{1}{\rho} \frac{\partial \tilde \psi}{\partial \theta} + \frac{1}{\rho} \frac{\partial f_{eq}}{\partial \theta} \frac{\partial \tilde \psi}{\partial \rho} \right)
\end{equation}
\begin{equation}
\label{eq:curv}
\sqrt{g} \left( \nabla \sqrt{g} \times \nabla \tilde f \right)^\zeta = \frac{\partial \sqrt{g} }{\partial \rho}  \frac{1}{\rho} \frac{\partial \tilde f}{\partial \theta} - \frac{1}{\rho} \frac{\partial \sqrt{g} }{\partial \theta} \frac{\partial \tilde f}{\partial \rho}
\end{equation}
with the Jacobian of the transformation:
\begin{equation}
\label{eq:Jac}
\frac{1}{\sqrt{g}} = \frac{B^2}{\varepsilon^2 (J+ \rlap{-} \iota I)}
\end{equation}

The geodesic compressibility in the frequency range of the geodesic acoustic mode (GAM) is included in the model by the parallel momentum response of the thermal plasma in the equations~\ref{pressure} and~\ref{velthermal} \cite{80,81}.

The equations are written using the equilibrium flux coordinates  $(\rho, \theta, \zeta)$ with $\rho$ the generalized radial coordinate proportional to the square root of the toroidal flux function (normalized to one at the edge) and $\theta$ the poloidal angle. We use the Boozer formulation for the flux coordinates \cite{82} with $\sqrt g$ the Jacobian of the coordinate transformation. The functions have two components, equilibrium and perturbation, represented as: $ A = A_{eq} + \tilde{A} $.

The FAR3D code uses finite differences in the radial direction and Fourier expansions in the two angular variables. The numerical scheme is semi-implicit in the linear terms. The nonlinear version uses a two semi-step method to ensure $(\Delta t)^2$ accuracy.

Many of the results in this paper have been obtained utilizing an eigenvalue solver, which is a recently added feature to the FAR3D model. In the linear regime, the time variation of all quantities will asymptote to $e^{-i \omega t}$; in this limit the equations of the gyrofluid model can be
cast into the form of a generalized eigenvalue problem: $\omega Ax = Bx$. This problem can be solved using an extension of the Jacobi-Davidson algorithm that was applied in \cite{83} for stable shear Alfv\'en eigenmodes to complex eigenvalues. For the complex case the user provides a target value for the frequency and growth rate and the algorithm returns a number of eigenvalues that are near the target in the complex plane. This method is especially useful in searching for higher frequency Alfv\' enic instabilities when lower frequency MHD instabilities are present with higher growth rates. If the initial value approach was used, only the faster growing MHD modes would be visible. The identification of subdominant and damped modes is also becoming an important tool \cite{84} in understanding the nonlinear dynamics of plasma instabilities.

The model was validated on previous studies of AE activity in LHD \cite{85,86}, TJ-II \cite{87,88,89} and DIII-D \cite{90,91} indicating a reasonable agreement with the experimental observations, identifying the main instability properties of the discharges, that is to say, the instabilities associated with the dominant modes and most important sub-dominant modes. It should be noted that the destabilizing effect of the thermal ion temperature gradients are not included in the model, so the Alfvenic ion temperature gradient (AITG) instability is out of the scope of the present analysis \cite{92,93}. The only thermal plasma drive included here is through the total thermal plasma pressure gradient. However, this only provides a drive for MHD ballooning modes, not for ITG instabilities. A global gyrofluid model has been constructed for ITG modes \cite{94} and, in principle, such drives could be included in this model, but this will remain a topic for future research.

The EP and thermal ion finite Larmor radius damping effects as well as the electron-ion Landau damping effects are not included in the study for simplicity, leading to simulations with larger growth rates as compared with the experimental observations. Nevertheless, the simulations reproduce the main instabilities observed in the experimental data. Also, an analysis based on calculations using the instability growth rates is more robust from the point of view of the optimization as discussed in section \ref{sec:Opt}.

\subsection{Equilibrium properties}

The table~\ref{Table:1} shows the shot numbers and time frames of the fixed boundary VMEC equilibrium used in the simulations, calculated from the DIII-D reconstruction of reverse magnetic shear discharges \cite{77}.

\begin{table*}[t]
\centering
\begin{tabular}{c}
Shot $164841$ (Low frequency AEs)
\end{tabular}

\begin{tabular}{c | c c c c c c c }
Time (ms) & $n_{i}(0)$ ($10^{20}$ m$^{-3}$) & $T_{i}(0)$ (keV) & $\beta_{th}(0)$ & $n_{f}(0)$ ($10^{20}$ m$^{-3}$) & $T_{f}(0)$ (keV) & $\beta_{f}(0)$ & $\omega_{cy}\tau_{A0}$\\ \hline
1740 (M10) & 0.28 & 4.66 & 0.059 & 0.047 & 33.1 & 0.0179 & 41.54 \\
2400 (M1A) & 0.29 & 5.11 & 0.079 & 0.062 & 36.7 & 0.0263 & 42.00 \\
2560 (M1B) & 0.29 & 5.46 & 0.056 & 0.066 & 38.5 & 0.0293 & 42.11 \\
2700 (M1C) & 0.29 & 5.54 & 0.086 & 0.064 & 37.9 & 0.0280 & 42.73 \\
\end{tabular}

\begin{tabular}{c}
Shot $164842$ (Core positive shear)
\end{tabular}

\begin{tabular}{c | c c c c c c c }
Time (ms) & $n_{i}(0)$ ($10^{20}$ m$^{-3}$) & $T_{i}(0)$ (keV) & $\beta_{th}(0)$ & $n_{f}(0)$ ($10^{20}$ m$^{-3}$) & $T_{f}(0)$ (keV) & $\beta_{f}(0)$ & $\omega_{cy}\tau_{A0}$\\ \hline
1580 (M20) & 0.17 & 4.74 & 0.064 & 0.005 & 33.3 & 0.0018 & 32.7 \\
2505 (M2A) & 0.18 & 5.15 & 0.076 & 0.057 & 35.7 & 0.0234 & 33.8 \\
2705 (M2B) & 0.15 & 5.15 & 0.092 & 0.091 & 35.4 & 0.0372 & 29.91 \\
2905 (M2C) & 0.13 & 5.15 & 0.097 & 0.104 & 35.4 & 0.0479 & 29.12 \\
\end{tabular}

\begin{tabular}{c}
Shot $164922$ (Core negative shear)
\end{tabular}

\begin{tabular}{c | c c c c c c c }
Time (ms) & $n_{i}(0)$ ($10^{20}$ m$^{-3}$) & $T_{i}(0)$ (keV) & $\beta_{th}(0)$ & $n_{f}(0)$ ($10^{20}$ m$^{-3}$) & $T_{f}(0)$ (keV) & $\beta_{f}(0)$ & $\omega_{cy}\tau_{A0}$\\ \hline
1600 (M30) & 0.28 & 5.02 & 0.075 & 0.101 & 33.4 & 0.0388 & 59.37 \\
2505 (M3A) & 0.35 & 5.93 & 0.103 & 0.083 & 37.9 & 0.0363 & 47.4 \\
2705 (M3B) & 0.40 & 6.58 & 0.118 & 0.075 & 42.2 & 0.0364 & 50.5 \\
2905 (M3C) & 0.39 & 6.57 & 0.110 & 0.064 & 43.1 & 0.0317 & 50.0 \\
\end{tabular}

\caption{Plasma parameters (at the magnetic axis) at different times along the discharges. First column is the thermal ion density, second column is the thermal ion temperature, third column is the thermal $\beta$, forth column is the energetic particle density, fifth column is the energetic particle temperature, sixth column the energetic particle $\beta$ and seventh column the normalized cyclotron frequency.} \label{Table:1}
\end{table*}

The shot $164841$ is a case with strong low frequency AE activity, shot $164842$ a discharge with central positive shear and shot $164922$ a discharge with central negative shear. The equilibrium reconstruction is done using the following experimental constrains: magnetic data, MSE data, kinetic pressure and edge density profile from NEO model. We use up-down symmetric equilibria, although deviations from the original single-null divertor equilibria are small. We analyze four different phases during the discharges: phase 0 before the formation of the reverse shear region, phase A during the destabilization of AE activity with steady frequencies, phase B during the destabilization of AE activity with non steady frequencies and phase C before the ramp down. Table~\ref{Table:1} summarizes the main plasma parameters. The magnetic field at the magnetic axis is $1.87$ T, the averaged inverse aspect ratio is $\varepsilon=0.48$ and the average elongation is $\Delta = 1.7$. 

\begin{figure}[h!]
\centering
\includegraphics[width=0.3\textwidth]{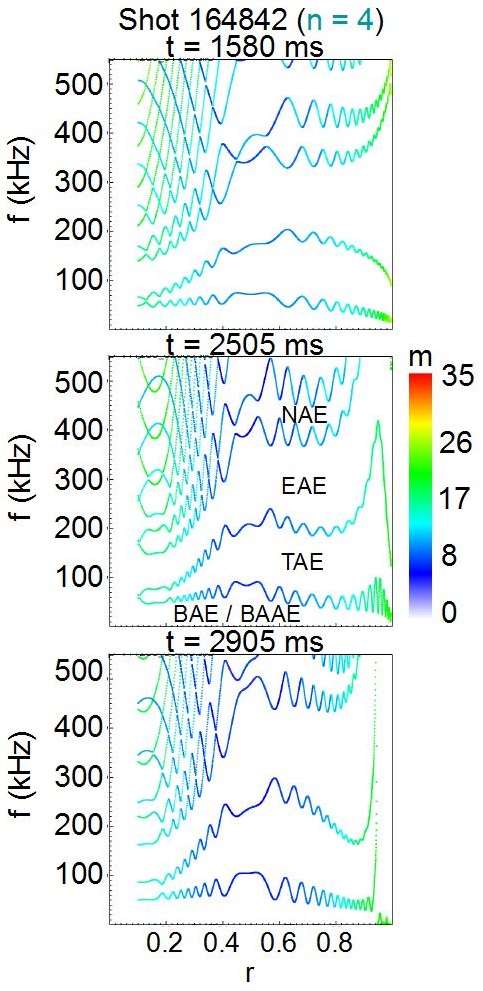}
\caption{Alfv\' en gaps of the $n=4$ toroidal mode of shot $164842$ at the different discharge phases.}\label{FIG:1}
\end{figure}

Figure~\ref{FIG:1} shows the Alfv\' en gaps of $n=4$ toroidal mode of shot $164842$ at the different discharge phases. There are four main Alfv\' en gaps: below $50$ kHz (BAE, BAAE and GAE gap), between $[50,175]$ kHz (TAEs gap), between $[175,350]$ kHz (EAE gap) and above $350$ kHz (NAE gap). The other shots analyzed in present study show similar gap distribution. 

\subsection{Simulation parameters}

Table~\ref{Table:2} shows the dynamic and equilibrium toroidal (n) and poloidal (m) modes for each model. The plasma edge ($\rho < 0.75$) is not included in the analysis of M2 and M3 cases to avoid he excitation of artificial ideal MHD instabilities at the outer boundary. The mode number notation is $m/n$ consistent with the $q=m/n$ definition. The simulations have a uniform radial grid of 1000 points (200 points for the subdominant modes studies).

\begin{table*}[h]
\centering

\begin{tabular}{c}
Dynamic (m) modes
\end{tabular}

\begin{tabular}{c | c c c c c c}
\hline
Case & $n=1$ & $n=2$ & $n=3$ & $n=4$ & $n=5$ & $n=6$ \\ \hline
M10 & $[3,5]$ & $[5,8]$ & $[8,12]$ & $[10,16]$ & $[13,20]$ & $[15,24]$ \\
M1A & $[1,3]$ & $[3,6]$ & $[5,10]$ & $[7,13]$ & $[9,16]$ & $[10,19]$ \\
M1B & $[2,4]$ & $[3,6]$ & $[5,10]$ & $[6,13]$ & $[8,16]$ & $[10,19]$ \\
M1C & $[1,3]$ & $[3,6]$ & $[5,10]$ & $[6,13]$ & $[8,16]$ & $[10,19]$ \\ 
M20 & $[3,4]$ & $[6,8]$ & $[9,12]$ & $[12,16]$ & $[14,20]$ & $[17,24]$ \\
M2A & $[3,4]$ & $[5,8]$ & $[7,12]$ & $[10,16]$ & $[12,20]$ & $[14,24]$ \\
M2B & $[3,4]$ & $[5,8]$ & $[7,12]$ & $[10,16]$ & $[12,20]$ & $[14,24]$ \\
M2C & $[2,4]$ & $[4,8]$ & $[6,12]$ & $[8,16]$ & $[10,20]$ & $[12,24]$ \\ 
M30 & $[2,4]$ & $[4,8]$ & $[6,12]$ & $[8,16]$ & $[11,20]$ & $[12,24]$ \\
M3A & $[1,4]$ & $[3,8]$ & $[5,12]$ & $[7,16]$ & $[9,18]$ & $[10,19]$ \\
M3B & $[1,4]$ & $[3,8]$ & $[5,12]$ & $[7,16]$ & $[9,18]$ & $[10,19]$ \\
M3C & $[1,4]$ & $[3,8]$ & $[4,12]$ & $[6,14]$ & $[8,16]$ & $[8,16]$ \\ \hline
\end{tabular}

\begin{tabular}{c}
Equilibrium (m) modes
\end{tabular}

\begin{tabular}{c | c c c c c c}
\hline
Case & $n=1$ & $n=2$ & $n=3$ & $n=4$ & $n=5$ & $n=6$ \\ \hline
All & $[0,9]$ & -- & -- & -- & -- & -- \\ \hline
\end{tabular}

\caption{Dynamic and equilibrium toroidal (n) and poloidal (m) modes.} \label{Table:2}
\end{table*}

The MHD parities are broken by the kinetic closure moment equations (6) and (7) so both parities are included in the study. For all the dynamic variables $sin(m\theta + n\zeta)$ and $cos(m\theta + n\zeta)$ parities are considered in the calculation of the growth rate and real frequency. The convention is as follows: $n > 0$ corresponds to the Fourier component $\cos(m\theta + n\zeta)$ and $n < 0$ to $\sin(-m\theta - n\zeta)$ (pressure eigenfunction case). The magnetic Lundquist number is $S=5\cdot 10^6$ similar to the experimental value in the middle of the plasma.

In the simulations the NBI injection intensity is controlled by fast ion $\beta$ value ($\beta_{f}$), linked to the density ratio between energetic particles and bulk plasma $n_{f}(0)/n_{e}(0)$ at the magnetic axis, calculated by the code TRANSP without the effect of the anomalous beam ion transport. The beam energy and the resonance regime between thermal plasma and energetic particles is controlled by the ratio of the energetic particle thermal velocity to the Alfv\' en velocity at the magnetic axis ($v_{th,f}/v_{A0}$). The energetic particle distribution function is a Maxwellian.  

The shots analyzed use co-injected NBI. These discharges are categorized as negative shear cases although during the shot $164922$ there are phases where the safety factor shows a positive shear near the magnetic axis, reason why we identify this discharge as a positive shear case \cite{95}.

\section{Dominant and subdominant MHD modes and AE \label{sec:main}}

In this section we analyze the growth rate ($\gamma$) and frequency ($f$) of the dominant and sub-dominant MHD and AE modes. We compare the simulation results with the instabilities observed in the experiments by magnetic and CO2 interferometry diagnostics. The instabilities with the largest growth rate that can constrain the device performance are identified in each phase of the discharge.

\subsection*{Low frequency AEs}

\begin{figure}[h!]
\centering
\includegraphics[width=0.45\textwidth]{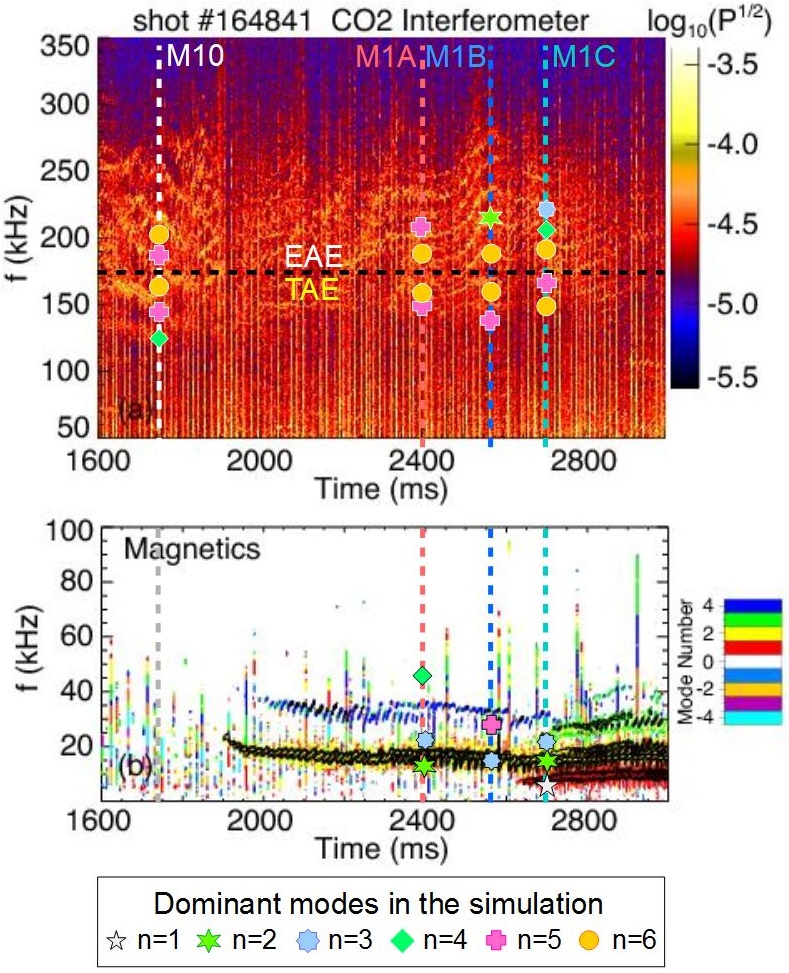}
\caption{Instabilities measured in the shot $164841$ at different phase during the discharge: M10 ($t=1740$ ms, white dotted line), M1A ($t=2400$ ms, red dotted line), M1B ($t=1560$ ms, blue dotted line) and M1C ($t=2700$ ms, cyan dotted line). The colored stars indicate the dominant modes of the simulation. The black dashed line indicates the transition between TAE and EAE families.}\label{FIG:2}
\end{figure}

Figure~\ref{FIG:2} shows the instabilities measured during shot $164841$. In this discharge both CO2 interferometer and magnetic data are available. The CO2 interferometry shows AE activity throughout the discharge: burst activity at M10, constant frequency AE at M1A, up-sweeping frequency AE at M1B and weaker steady frequency AE at M1C. In addition, the magnetic diagnostic shows $n=1$ to $4$ low frequency instabilities in the range of $5$ to $40$ kHz at different phases of the discharge. In figure~\ref{FIG:2} the colored stars show frequencies of the dominant modes in the simulations overlaid on the experimental diagnostic signals. As can be seen the simulations bracket many of the experimentally measured frequency lines. However, they miss some of the higher frequency lines for the M10, M1A, and M1B time slices. Also, in some cases modes are predicted at lower frequencies (M10 and M1B) than seen experimentally.

\begin{figure}[h!]
\centering
\includegraphics[width=0.45\textwidth]{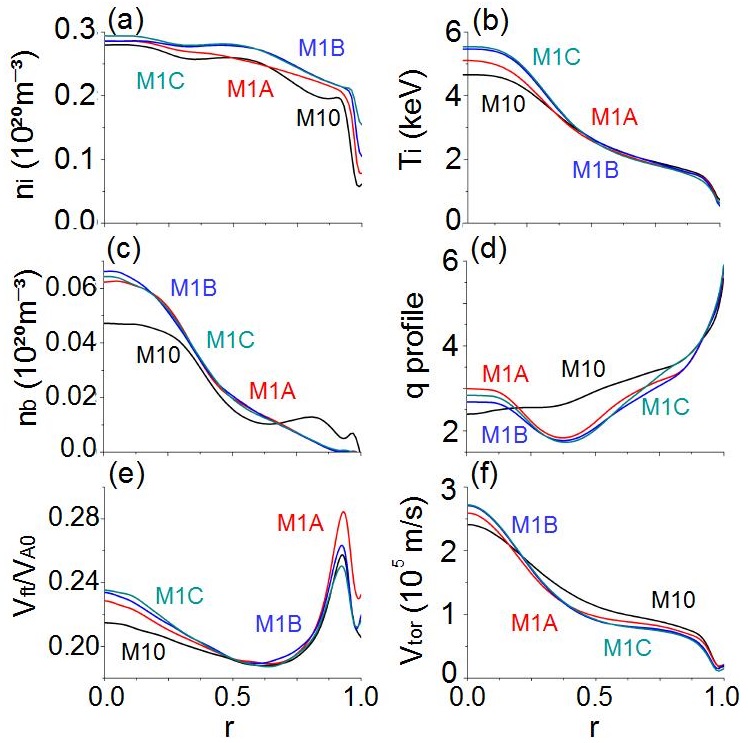}
\caption{Shot $164841$ (a) thermal ion density, (b) thermal ion temperature, (c) energetic particle density, (d) q profile, (e) $V_{th,f}/V_{A0}$ ratio and (f) toroidal rotation at different discharge phases: M10 (black), M1A (red), M1B (blue) and M1C (cyan).}\label{FIG:3}
\end{figure}

Figure~\ref{FIG:3} shows the plasma profiles and magnetic field configuration at different phases of the discharge. It should be noted that the energetic particle density gradient and the reverse shear region are both located between the inner and middle plasma, so RSAE are easily destabilized.

\begin{figure}[h!]
\centering
\includegraphics[width=0.5\textwidth]{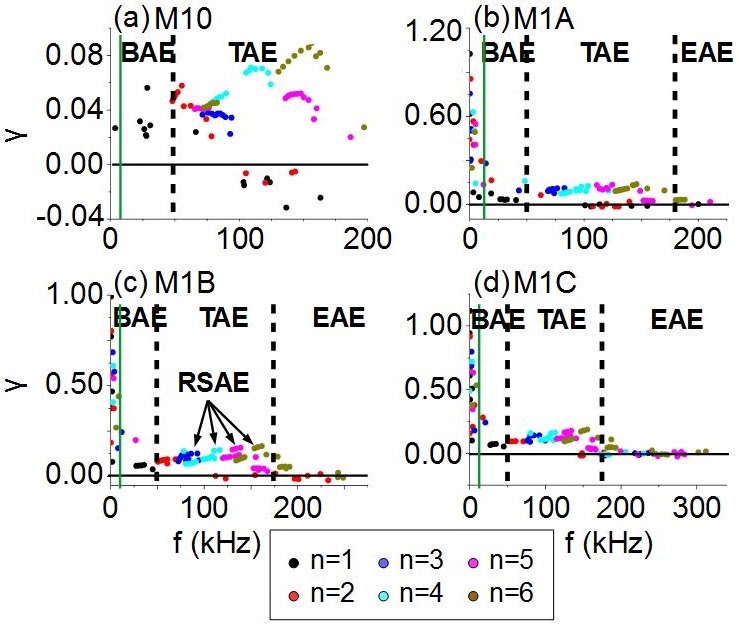}
\caption{Growth rate and frequency of the dominant and subdominant modes ($n=1$ to $6$) for $164841$ shot phases (a) M10, (b) M1A, (c) M1B and (c) M1C. The modes below the solid black line are stable damped modes. The solid green line separates MHD-like modes and Alfv\' en Eigenmodes. The dashed black lines separate different AE families (TAE/EAE/NAE).}\label{FIG:4}
\end{figure}

Figure~\ref{FIG:4} shows the growth rate and frequency of dominant and subdominant modes. At M10 phase, only n=1 modes are MHD unstable although all n modes are AE unstable: $n=1$ BAE with $f \approx 30$ kHz and $n=2$ to $n=6$ TAE with $f \approx 55$, $75$, $120$, $150$ and $160$ kHz. The AEs growth rate is between 2 to 3 times larger than that of the MHD modes. This result is consistent with the observation that the discharge shows strong AE activity in its early phase. At M1A phase, all modes are MHD unstable with growth rates 2 and 3 times larger than the unstable AE: $n=1$ and $n=2$ BAE with $f \approx 25$ kHz, $n=4$ BAE with $f \approx 45$ kHz, $n=3$ to $n=6$ TAE with $f \approx 75$, $105$, $125$ and $150$ kHz and $n=5$ and $6$ EAE with $f \approx 175$ and $195$ kHz. The strong $n=2$ instability (with the $n=4$ overtone) observed by the magnetic diagnostic is reproduced in the simulations as $n=2$ and $n=4$ BAE, as well as the $n=5$ and $n=6$ TAE/EAE, consistent with the range of instability frequencies measured by CO2 interferometer. The enhancement of the MHD activity is linked to the formation of the reverse shear region in the inner-middle plasma, as well as an increase of the AE growth rate by around $50\%$. At M1B phase, again the most unstable modes are MHD but the growth rate is slightly smaller compared to M1A phase, although the growth rate of the AEs is similar for the $n=1$ BAE with $f \approx 30$ kHz, the $n=2$ to $n=6$ TAE with $f \approx 60$, $95$, $125$, $145$ and $160$ kHz as well as $n=5$ and $6$ EAE with $f \approx 175$ and $190$ kHz, showing an increase in the AE frequency compared to M1A phase. The main characteristic of this discharge phase is the destabilization of RSAEs, resulting in frequencies in the range of the TAE gap and growth rates slightly larger than the TAEs, consistent with the up-sweeping frequency AEs observed in the experiment. The destabilization of the RSAE is caused by a slight increase of the EP density gradient and a decrease of the reverse shear region depth. At M1C phase, the most unstable modes are still MHD as well as the subdominant $n=2$ and $3$ BAE with $f \approx 25$ and $30$ kHz, similar to the magnetic diagnostic observations. In addition $n=1$ BAE with $f \approx 45$ kHz, $n=2$ to $n=6$ TAE with $f \approx 60$, $90$, $110$, $140$ and $155$ kHz as well as $n=5$ and $6$ EAE with $f \approx 175$ and $200$ kHz are unstable. It should be noted that the simulations identify instabilities in the frequency range of $f = [70,125]$ kHz not observed in the experimental data. The instabilities calculated in the simulations in that frequency range are TAE, RSAE and BAE destabilized between the middle and the outer plasma region where the Alfven gap is extended to higher/lower frequencies, as well as sub-dominant energetic particles modes (EPM) destabilized in the continuum. The reason these modes are unstable in the simulations but not observed in the experiment is because the model doesn't include the EP finite Larmor radius or electron-ion Landau damping effects, that should lead to the stabilization of these modes. Future studies will include these damping effects to improve the simulation results. In addition, the safety factor and energetic particle profiles are not directly measured during the experiment, affecting the AE stability threshold obtained in the simulations. Newer fast ion Dα (FIDA) measurements in DIII-D show that the action of AE modes is to flatten the EP profiles, thus the EP density profiles could be less peaked than the configuration analyzed.

In summary, DIII-D performance in the early phase of the M1 discharge is limited by the AE activity driven in the inner plasma region before the reverse shear is formed. After the formation of the reverse shear region, MHD-likes modes are dominant and coexist with subdominant AEs. During the late phase of the discharge, RSAE are destabilized when the EP density gradient reaches it maximum value inside the reverse shear region. The optimization of the discharge stability based on these calculations would require modified phasing of the NBI power, particularly at the beginning of the discharge, and the amelioration of the MHD stability as soon as the reverse shear region is formed. The reason why this discharge has a bad MHD stability is because the inner plasma core resonates with the $q=3$ rational surface and the bottom of the reverse shear region with $q=2$ rational surface.

A more detailed discussion of how the AE families are identified by analyzing the individual eigenfunctions is given in the Appendix, as well as a comparison with the experimental data.

\subsection*{Central negative shear}

\begin{figure}[h!]
\centering
\includegraphics[width=0.45\textwidth]{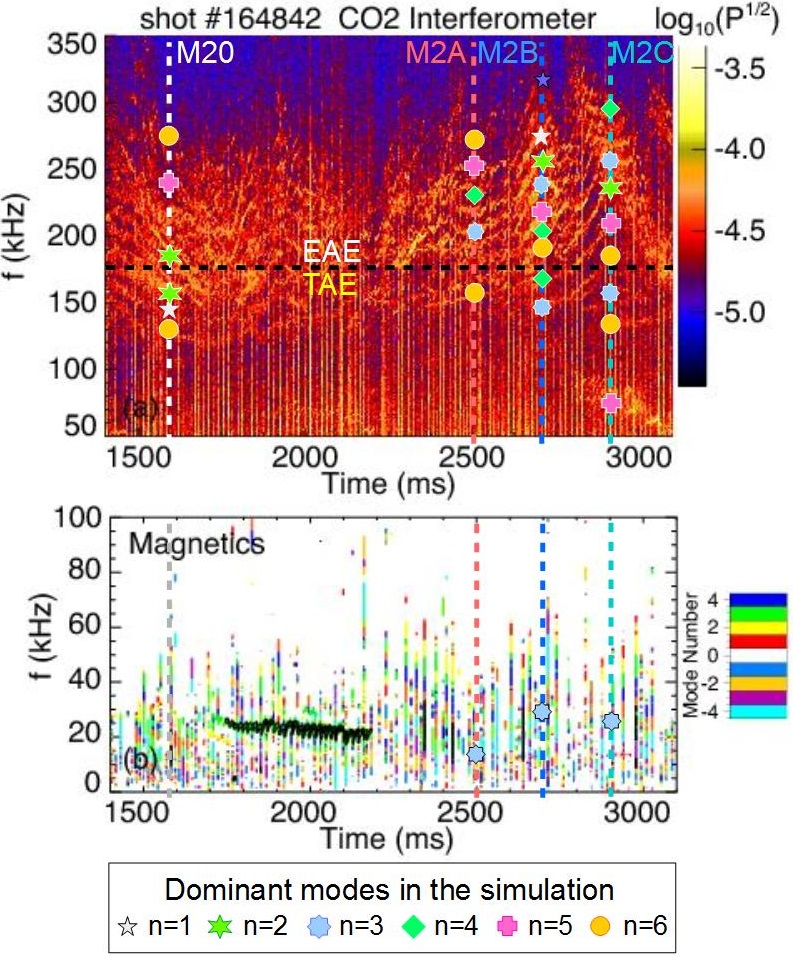}
\caption{Instabilities measured in the shot $164842$ at different phase during the discharge: M20 ($t=1580$ ms, white dotted line), M2A ($t=2505$ ms, red dotted line), M2B ($t=2705$ ms, blue dotted line) and M2C ($t=2905$ ms, cyan dotted line). The colored stars indicate the dominant modes of the simulation. The black dashed line indicates the transition between TAE and EAE families.}\label{FIG:5}
\end{figure}

Figure~\ref{FIG:5} shows the instabilities measured during shot $164842$. CO2 interferometry shows AE instabilities during all the discharge: burst activity at M20, constant frequency AE at M2A, up-sweeping frequency AE at M2B and down-sweeping frequency AE at M2C. The frequencies of the dominant modes in the simulations are consistent with the experimentally measured frequency lines, although some higher frequency lines are missing.

\begin{figure}[h!]
\centering
\includegraphics[width=0.45\textwidth]{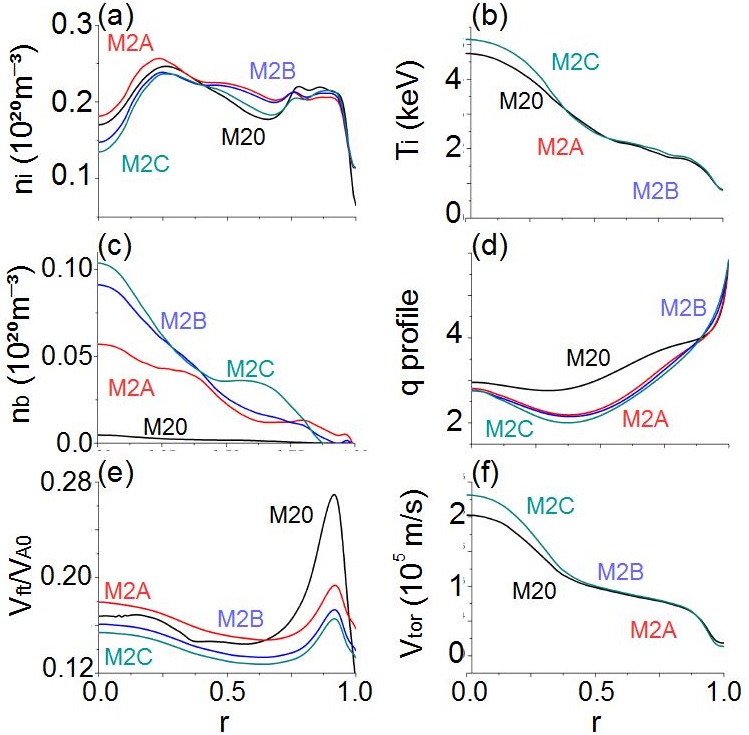}
\caption{Shot $164842$ (a) thermal ion density, (b) thermal ion temperature, (c) energetic particle density, (d) q profile, (e) $V_{th,f}/V_{A0}$ ratio and (f) toroidal rotation at different discharge phases: M20 (black), M2A (red), M2B (blue) and M2C (cyan).}\label{FIG:6}
\end{figure}

Figure~\ref{FIG:6} shows the plasma profiles at different phases of the discharge. The configuration is similar to the $164841$ shot although the $q$ profile near the magnetic axis is non constant and there is a negative shear region in the core.

\begin{figure}[h!]
\centering
\includegraphics[width=0.5\textwidth]{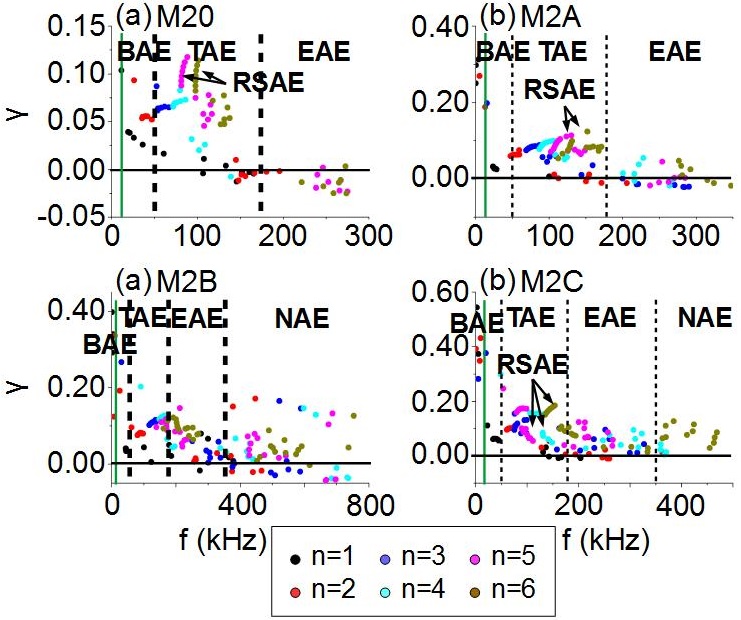}
\caption{Growth rate and frequency of dominant and subdominant modes ($n=1$ to $6$) for $164842$ shot phases (a) M20, (b) M2A, (c) M2B and (c) M2C. The modes below the solid black line are stable damped modes. The solid green line separates MHD-like modes and Alfv\' en Eigenmodes. The dashed black lines separates different AE families (TAE/EAE/NAE).}\label{FIG:7}
\end{figure}

Figure~\ref{FIG:7} shows the growth rate and frequency of dominant and subdominant modes at different phases of the discharge. The phases M20 and M2A show a similar AE stability than the M1 discharge, although the MHD stability is improved with only $n=1$ to $n=2$ unstable MHD modes, even if the reverse shear region was already formed at the M20 phase. On the other hand, $n=5$ and $n=6$ RSAE are destabilized during both phases. At phase M2B, $n=1$ to $n=3$ modes are MHD unstable as well as TAE/EAE/NAE of all toroidal modes in the range of frequencies $f = [60,700]$ kHz. The AE instabilities are stronger compared to the M1 discharge and the previous stages of the shot because there is a strong gradient of the EP density that covers the inner and middle plasma region, leading to the destabilization of high frequency AEs near the magnetic axis. At M2C phase, $n=1$ to $3$ MHD modes are unstable but the AE growth rates and frequencies are smaller compared to M2B phase because the EP density gradient only covers the inner plasma region.

Consequently, the MHD activity of shot $164842$ is smaller compared to $164841$ shot because the $q=3$ rational surface is non resonant in the inner plasma neither is $q=2$ near the bottom of the reverse shear region. On the other hand, AEs are destabilized from the first stage of the discharge, particularly the RSAE, because the energetic particle density gradient is located inside the reverse shear region. The discharge MHD and AE stability can be further improved increasing the magnetic shear in the inner plasma region, avoiding the destabilization of MHD modes and AE near the magnetic axis. In addition, unstable RSAE can be avoided if the energetic particle density gradient is not located near the bottom of the reverse shear region. Phase M2B shows the largest AEs growth rates and frequencies because the NBI beam energy and injection intensity are large enough to destabilize EAE and NAE modes of several toroidal families in the inner plasma region.

\subsection*{Central positive shear}

\begin{figure}[h!]
\centering
\includegraphics[width=0.45\textwidth]{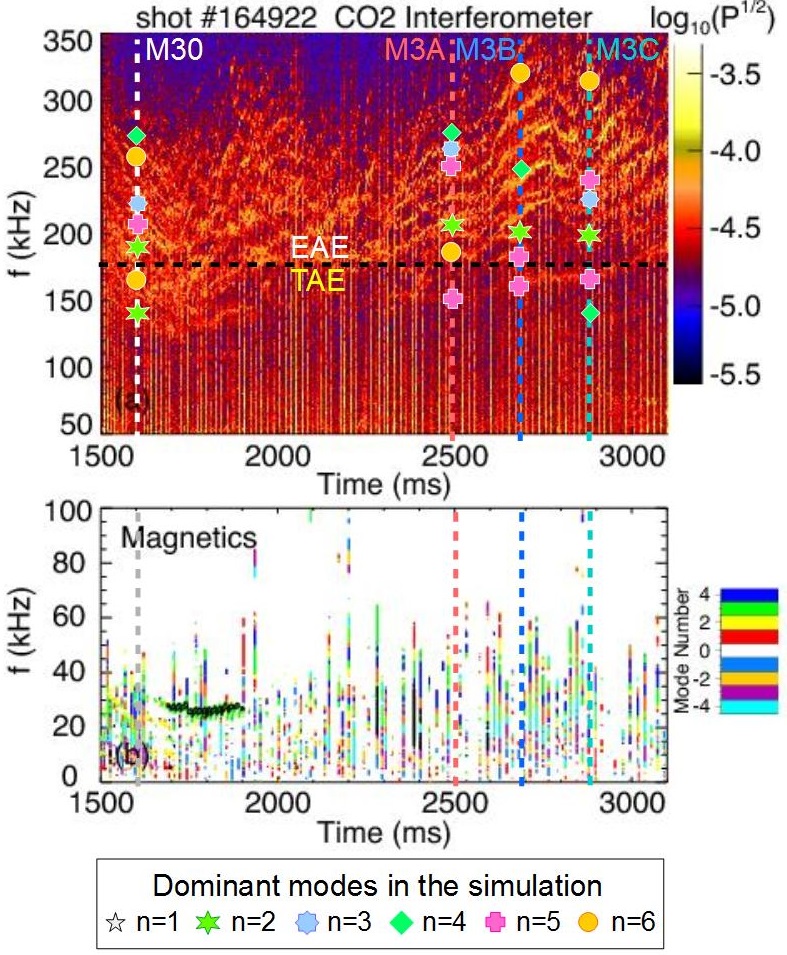}
\caption{Instabilities measured in the shot $164922$ at different phase during the discharge: M30 ($t=1600$ ms, white dotted line), M3A ($t=2500$ ms, red dotted line), M3B ($t=2700$ ms, blue dotted line) and M3C ($t=2900$ ms, cyan dotted line). The colored stars indicate the dominant modes of the simulation. The black dashed line indicates the transition between TAE and EAE families.}\label{FIG:8}
\end{figure}

Figure~\ref{FIG:8} shows the instabilities measured during shot $164922$. CO2 interferometry shows AE instabilities throughout the discharge: down-sweeping frequency AE at M30, constant frequency AE at M3A, up-sweeping frequency AE at M3B and high constant frequency AE ($f > 250$ kHz) at M3C. Again, the frequencies of the dominant modes in the simulations are similar to the experimentally measured frequency lines, missing some higher frequency lines particularly in the M3B and M3C discharge phases.

\begin{figure}[h!]
\centering
\includegraphics[width=0.45\textwidth]{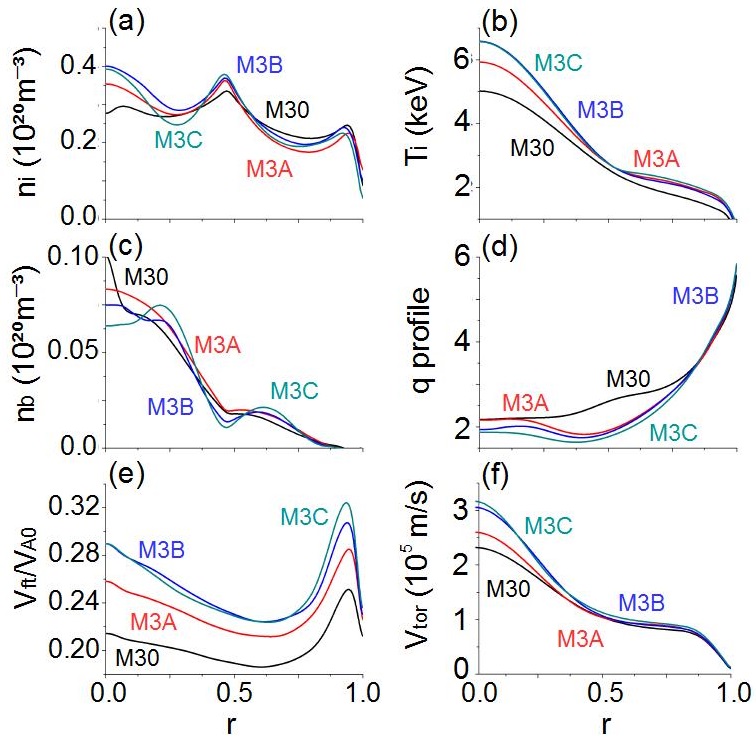}
\caption{Shot $164922$ (a) thermal ion density, (b) thermal ion temperature, (c) energetic particle density, (d) q profile, (e) $V_{th,f}/V_{A0}$ ratio and (f) toroidal rotation at different discharge phases: M30 (black), M3A (red), M3B (blue) and M3C (cyan).}\label{FIG:9}
\end{figure}

Figure~\ref{FIG:9} shows the plasma profiles. The averaged thermal ion temperature and density are larger compared to shot $164842$. The energetic particle density gradient is narrower and located in the inner plasma region (between $0.2 < r < 0.4$), the $V_{th,f}/V_{A0}$ ratio is larger and there is a region of negative shear in the plasma core.

\begin{figure}[h!]
\centering
\includegraphics[width=0.5\textwidth]{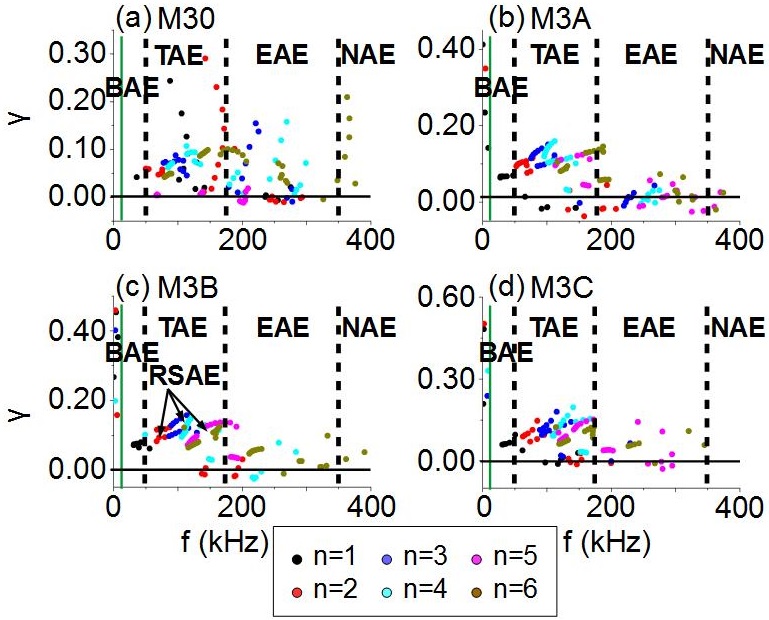}
\caption{Growth rate and frequency of the dominant and subdominant modes ($n=1$ to $6$) for $164922$ shot phases (a) M30, (b) M3A, (c) M3B and (c) M3C. The modes below the solid black line are stable damped modes. The solid green line separates MHD-like modes and Alfv\' en Eigenmodes. The dashed black lines separates different AE families (TAE/EAE/NAE).}\label{FIG:10}
\end{figure}

Figure~\ref{FIG:10} shows the growth rate and frequency of dominant and subdominant modes. At phase M30, the plasma is MHD stable but several AEs are destabilized, particularly $n=1-2$ TAEs and $n=6$ NAE with large growth rates and $f \approx 95$, $145$ and $375$ kHz destabilized near the magnetic axis where the q profile is almost flat. Again, this indicates the tendency for a weak magnetic shear to deteriorate the AE stability of the plasma. At phase M3A, $n=1$ and $2$ MHD modes are destabilized with growth rates 2 times larger compared to the AEs, especially the TAEs show the largest growth rates while the NAE are marginally unstable because the EP density gradient is weaker close to the magnetic axis. At phase M3B, $n=1$ to $n=4$ modes are MHD unstable, as well as several subdominant AE, in particular RSAE are destabilized in the middle plasma region. Phases M3B and M3C show similar MHD and AE stability properties. The range of frequencies observed by the magnetic measurements of low frequency AE and CO2 interferometer data are consistent with the simulation results.  

These simulations indicate that the performance at the first stage of the discharge, while the reverse shear region is not yet formed, is limited by the TAE and EAE destabilized near the magnetic axis . During the rest of the discharge, $n=1$ to $n=4$ interchange modes are unstable in the bottom of the reverse shear region driven by $q=2$ rational surface. The magnetic shear near the inner plasma region is almost flat during all the discharge phases except at M3B phase, where TAE/RSAE are destabilized in the plasma core and in the bottom of the reverse shear region. In addition, subdominant TAE, EAE and NAE are destabilized in a range of frequencies up to $400$ kHz. BAE and low frequency TAE reproduced in the simulation are also consistent with the magnetic diagnostics, showing instabilities in the range of $f < 60 $ kHz during all the discharge phases.

In summary, after identifying the main dominant and sub-dominant modes that can impact the DIII-D heating efficiency at the different phases of the discharges, as well as some possible optimization trends to improve the plasma MHD and AE stability, in the next section we analyze the effect of the NBI operational regimen and magnetic field configuration on the plasma stability.

\section{Optimization trends to improve the MHD and AE stability \label{sec:Opt}}

In this section, based on the previous studies, we identify some optimization trends to improve the AE and MHD stability of the plasma by modifying the NBI operational regime and the magnetic field configuration.

\subsection{NBI operational regime}

We perform a parametric analysis varying the injection intensity ($\beta_{f}$), beam energy ($V_{th,f}/V_{A0}$) and the location of the EP density gradient ($r_{peak}$) to find the optimal operational regime of the NBI that minimizes AE activity. To study the effect of the EP density gradient location on the instabilities growth rate and frequency we define the next analytic expression for the EP density:

$$ n_{f}(r) = \frac{(0.5 (1+ \tanh(r_{flat}*(r_{peak}-r))+0.02)}{(0.5 (1+\tanh(r_{flat}*r_{peak}))+0.02)}$$
where $r_{flat}$ parameter controls the slope flatness of the profile (fixed to $r_{flat}=7$). The study is performed at phase B of the M2 and M3 discharges.

\begin{figure*}[h!]
\centering
\includegraphics[width=1.0\textwidth]{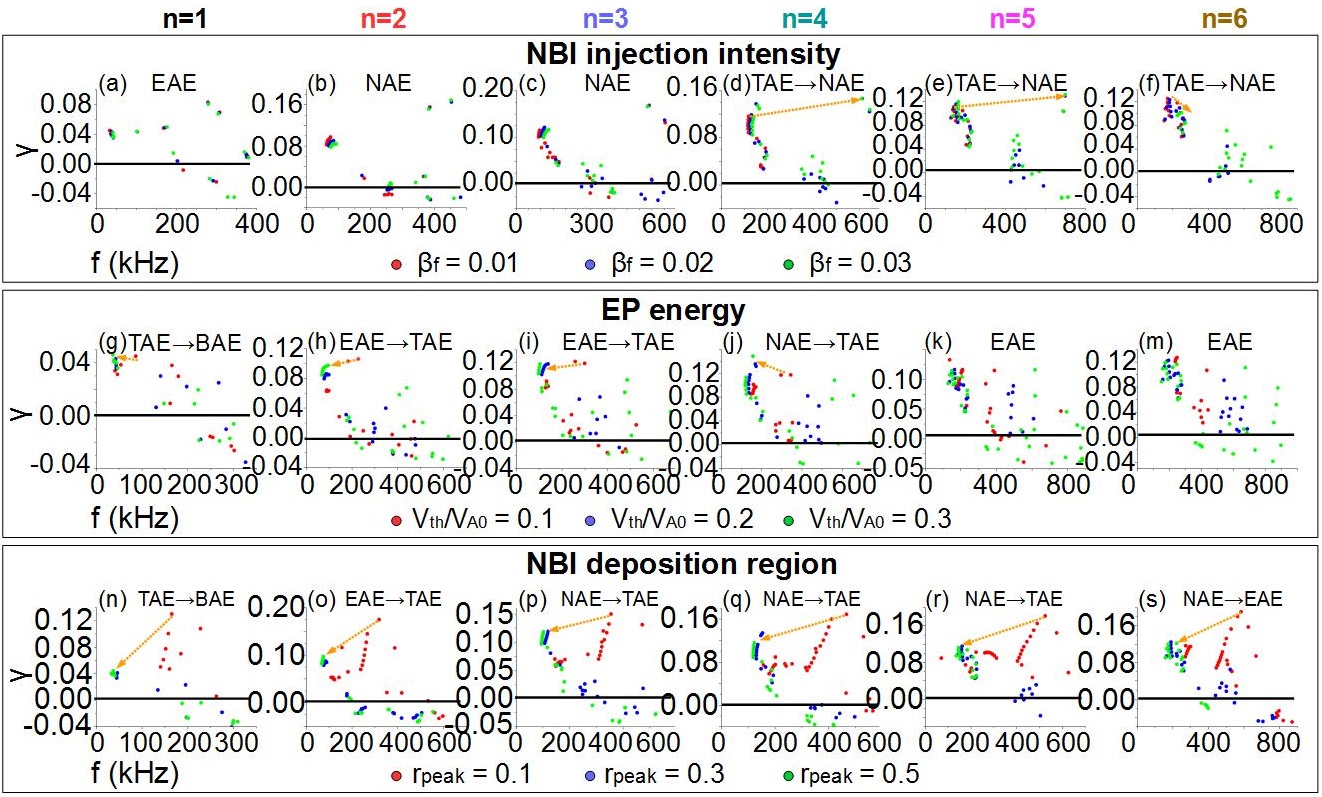}
\caption{Dependency of the growth rate and frequency of dominant and subdominant modes ($n=1$ to $6$) with the NBI injection intensity (panels a to f), beam energy (panels g to m) and deposition region (panels n to s) for M2B phase. The dominant AE family is indicated in the top of the panel and, if it takes place, the transition between different AE families (highlighted by a dotted orange arrow).}\label{FIG:11}
\end{figure*}    

Figure~\ref{FIG:11} shows the dependencies of growth rate and frequency for dominant and subdominant modes at different NBI injection intensities, beam energies and deposition regions for the M2B phase. Increasing $\beta_{f}$ from $0.01$ to $0.03$ (panels a to f) only leads to a transition between AE families (TAE $\rightarrow$ NAE, highlighted by a dotted orange arrow) for the $n=4$ to $6$ modes, although the critical $\beta_{f}$ to destabilize $n=1$ EAE and $n=2-3$ NAE is smaller than $0.01$, so there is only a partial optimization of the discharge AE stability if the NBI injection intensity is smaller than $\beta_{f} = 0.01$. Increasing the beam energy ($V_{th,f}/V_{A0}$ ratio, panels g to m) leads to a transition between AE families for $n=1$ to $n=4$ modes: from $n=1$ TAE to $n=1$ BAE, $n=2$ EAE to $n=2$ TAE, $n=3$ EAE to $n=3$ TAE and $n=4$ NAE to $n=4$ TAE. On the other hand, no transition is observed for $n=5$ and $6$ EAE, only a decrease of the AE growth rate and frequency that can be explained as a weaker resonance between the EP and the thermal plasma. Changing the location of the EP density gradient from $r_{peak}=0.1$ (on-axis NBI injection) to $r_{peak}=0.3$ and $0.5$ (off-axis NBI injection) leads to the stabilization of $n=1$ TAE, $n=2$ EAE and $n=3$ to $6$ NAE  (panels n to s), although $n=1$ BAE, $n=2-5$ TAE and $n=6$ to $6$ EAE are destabilized with lower growth rates, so the AE stability of the plasma is optimized if the NBI is deposited off-axis.

\begin{figure*}[h!]
\centering
\includegraphics[width=1.0\textwidth]{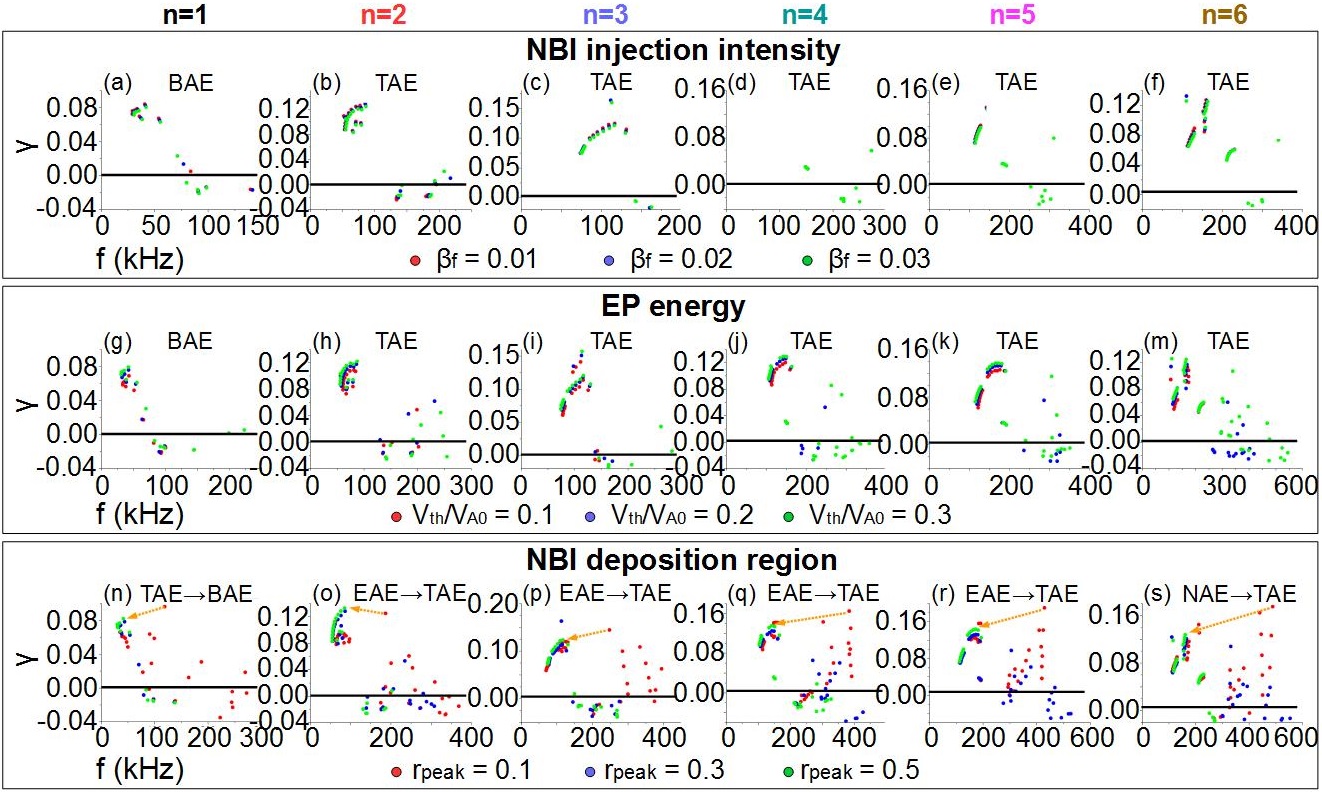}
\caption{Dependency of the growth rate and frequency of dominant and subdominant modes ($n=1$ to $6$) with the NBI injection intensity (panels a to f), beam energy (panels g to m) and deposition region (panels n to s) for M3B phase. The dominant AE family is indicated in the top of the panel and, if it takes place, the transition between different AE families (highlighted by a dotted orange arrow).}\label{FIG:12}
\end{figure*}    

Figure~\ref{FIG:12} shows the same analysis for the M3B phase. Modifying $\beta_{f}$ (panel a to f) or the beam energy (panels g to m) do not lead to the stabilization of AEs in this range of parameters; the growth rate and frequency are similar, so no optimization trends are observed. On the other hand, off axis NBI heating leads to stabilization of EAE/NAE (panels n to s).

In summary, an improved NBI operational scenario requires higher beam energy and off-axis NBI heating. Decrease the NBI injection energy doesn't lead to a significant improvement because the NBI operates above the critical $\beta_{f}$ value. It should be noted that there is experimental evidence of weaker AE activity if the beam energy decreases for the same injection intensity, as well as an enhancement of low frequency AEs. Consequently, an optimized NBI operational regime should not necessarily be linked to an arbitrary increase of the beam energy. In optimized NBI operational regimes, the EAE/NAE are stable and the growth rate/frequency of the BAE/TAE is smaller compared to the discharges analyzed. Previous studies also concluded that if the beam energy is high enough, the NBI operation entering into the non resonant regime where BAE/TAE modes are also stabilized \cite{84,87}.

\subsection{Magnetic field configuration}

\begin{figure}[h!]
\centering
\includegraphics[width=0.5\textwidth]{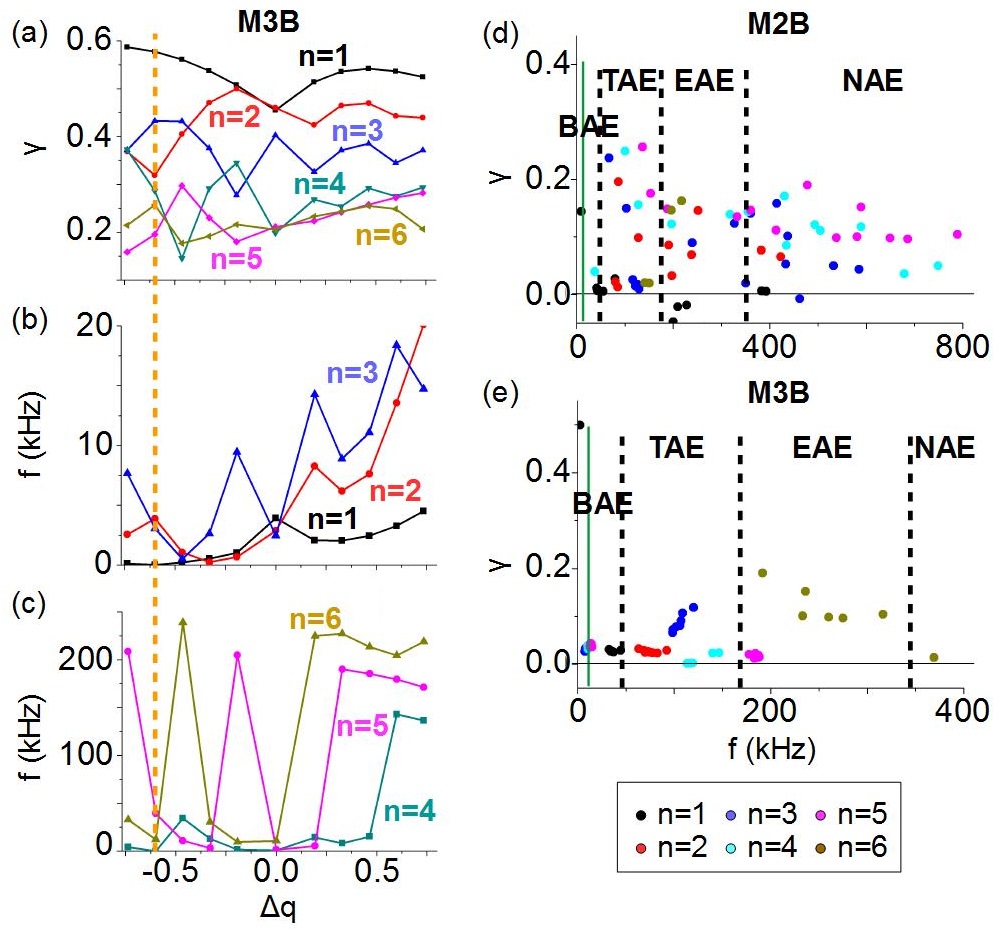}
\caption{Dependency of dominant modes growth rate and frequency ($n=1$ to $6$) with the magnetic field configuration. Panel (a) shows the modes growth rate versus $\Delta q$, (b) $n=1$ to $3$ modes frequency and (c) $n=4$ to $6$ modes frequency for case M3B (data for case 2B not shown). The dashed orange line indicates the optimized configuration selected to be further analyzed. Panel (d) shows the dominant and subdominant modes growth rate and frequency if the magnetic field configuration is modified for M2B ($\Delta q = -1$) and (e) M3B ($\Delta q=-0.6$)}\label{FIG:13}
\end{figure}

In this section, we analyze the effect of the magnetic field configuration with respect to MHD and AE stability at M2B and M3B phases. To accomplish this, the $q$ profile is displaced by $\Delta q = [-0.75,0.75]$ (Figure~\ref{FIG:13} case M3B) and we analyze the growth rate (panel a) and frequency (panels b and c) of the dominant $n=1$ to $6$ modes. $\Delta q > 0$ displacements lead to larger growth rate and frequency for all MHD and AE instabilities compared to $\Delta q < 0$ displacements. In addition, $\Delta q < 0$ displacements lead to operation windows with improved AE and MHD stability (for example $\Delta q = -0.6$ compared to $\Delta q = -0.465$).

To confirm the AE and MHD stability improvement we analyze the dominant and subdominant modes growth rate and frequency if the q profile is displaced $\Delta q = -1$ in M2B case and $\Delta q=-0.6$ in M3B case, see Figure~\ref{FIG:13} panels d) and e), corresponding to operation windows with optimized stability properties. The improved M2B case shows weaker MHD activity with only an $n=1$ mode being unstable at lower growth rate. In addition, BAEs are marginally unstable and the dominant modes are $n=2-5$ TAEs and $n=6$ EAE. The upgraded version of M3B also shows weaker MHD activity for all modes (except $n=1$, slightly larger), although the growth rate and frequency of the AEs is similar to the original cases.

Consequently, the improved M2B and M3B phases show better linear MHD and AE stability, so this optimization trend should be considered together with an improved NBI operational regime. Indeed, similar optimization trends were already observed by other authors identifying a reduction of the energetic particle transport in configurations with low $q_{min}$, although enhanced in configuration with high $q_{min}$ \cite{96,97,98,99}.

\section{Destabilizing effect of the energetic particles on MHD modes \label{sec:MHD/AE}}

\begin{figure}[h!]
\centering
\includegraphics[width=0.45\textwidth]{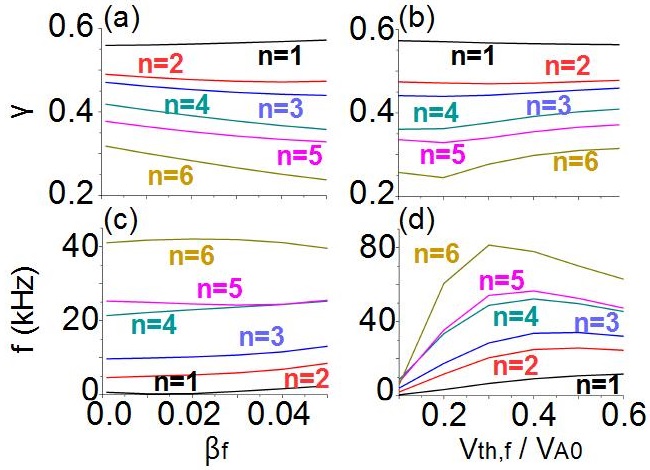}
\caption{Growth rate and frequency of the dominant modes ($n=1$ to $6$) versus $\beta_{f}$ (panels a and c) and $V_{th,f}/V_{A0}$ (panels b and d) for M2C phase.}\label{FIG:14}
\end{figure}

In this section we analyze the coexistence between MHD modes and low frequency AE, as well as the destabilizing effect of the energetic particles on MHD modes. If the MHD activity is dominant but the growth rate of the sub-dominant low frequency AE (BAE/BAAE) is not negligible (1/2 to 1/3 that of the dominant MHD modes), the MHD modes are further destabilized by the energetic particle drive so the MHD mode growth rate and frequency also change with the NBI injection intensity and beam energy (see Figure~\ref{FIG:14}, panels a) and c), modes $n=1$ and $n=2$). There is a smooth transition in terms of growth rate and frequency between MHD modes and low frequency AEs as the NBI injection intensity increases, pointing out that both instabilities should be partially overlapped and coexist in a range of $\beta_{f}$ values. On the other hand, the growth rate and frequency of MHD/AE instabilities show a stronger dependency with the beam energy, leading to a distinct transition between MHD modes and AEs, see for example the evolution of the $n=2$ MHD mode if $V_{th,f}/V_{A0}= 0.1$ to a $n=2$ BAE if $V_{th,f}/V_{A0} = 0.5$ (figure~\ref{FIG:14}, panels b and d).

In summary, DIII-D reverse magnetic shear discharges show an overlapping and possible coupling between MHD and AE low frequency instabilities, because the source of free energy added by the energetic particle density gradient further destabilizes unstable MHD modes in addition to the kinetic resonant destabilization of AEs. It should be noted that the destabilizing effect of the energetic particles on MHD modes is weak if the MHD modes growth rate is large compared to low frequency AE growth rate, negligible if the dominant MHD mode growth rate is 3 times larger. In addition, if the NBI operational regime is optimized to minimize AE activity (operation below the critical $\beta_{f}$ or in the weakly resonant regime), the overlapping between MHD modes and low frequency AE is weaker, optimizing the device performance with respect to low frequency AEs.

\section{Conclusions and discussion \label{sec:conclusions}}

The simulation results of dominant and subdominant modes in reverse magnetic shear discharges in DIII-D are in reasonable agreement with measurements. MHD modes are unstable as soon as the reverse shear region appears in the plasma, further enhanced as the reverse shear region deepens. In addition, unstable AEs may limit the device performance at the beginning of the discharge, before the formation of the reverse shear region, remaining unstable as subdominant modes through the rest of the discharge.

MHD modes are destabilized in the bottom of the reverse shear region near the $q=2$ rational surface, although if the $q=3$ rational surface is also resonant in the inner plasma region, as in the shot $164841$, MHD modes are destabilized close to the magnetic axis. AEs are unstable near the magnetic axis in the discharge with positive/negative magnetic shear at the plasma core, shots $164842$ and $164922$, because the magnetic shear at the plasma core is weak in both discharges. Core negative shear configurations can have a larger magnetic shear compared to core positive shear configurations avoiding the presence of main rational surfaces in the plasma core and in the reverse shear region, improving MHD and AE stability.

Thanks to the analysis of the subdominant modes in the simulations we can study the stability of different AE families as overtones of the same perturbation, reproducing the wide spectrum of AE frequencies observed in the experiment. For example, the numerical model predicts unstable AEs in the frequency range of $f = [125,350]$ kHz during the phases B and C of shot $164842$, reproducing the CO2 interferometry measurements. The magnetic diagnostic shows the destabilization of low frequency AE, TAE and MHD modes that coexist in a frequency range between $f =  [5,80]$ kHz. It should be noted that some discrepancies between the experimental data and the simulations are identified, for example the destabilization of modes in the frequency range of $f = 70$ to $125$ kHz not observed in the experiment. Such a mismatch can be explained by the stabilizing effect of the EP finite Larmor radius or electron-ion Landau damping, not considered in the study. Also, the fact that the safety factor and energetic particle profiles are not directly measured during the experiment, and the profiles can sensitively affect the AE stability threshold.

The NBI operational regime can be optimized to reduce the AE activity if the NBI deposition is off-axis and the beam energy increases (compared to the experimental NBI operational regime), reducing the AE frequency and growth rate and stabilizing EAE/NAE. If the beam energy is further increased, the NBI operation can enter in the non resonant regime, where TAEs, low frequency AEs or RSAE are stable, as was predicted and observed in previous studies of AE stability in LHD, TJ-II and DIII-D \cite{88,89,91}. It should be noted that reducing the beam energy decreases the number of lost EPs for the same injection intensity, so this optimization trend can't be ignored. On the other hand, the NBI injection intensity is above the critical $\beta_{f}$ to destabilize AEs, so no improvement is observed in the simulations even if $\beta_{f}$ is reduced down to $0.01$. The magnetic field configuration can also be optimized because there are operation windows with improved MHD and AE linear stability, particularly if the $q$ profile in the plasma core and reverse shear region is located between the $q=1$ and $q=2$ rational surfaces.

Energetic particle drive can further destabilize MHD modes whose growth rate and frequency change with the NBI injection intensity and beam energy. The simulations show an overlapping between MHD modes and low frequency AE as well as a smooth transition between MHD modes and AE if $\beta_{f}$ increases, so cross-perturbation effects should be expected. Nevertheless, MHD modes and low frequency AE overlapping can be avoided or minimized if the NBI operates below the critical $\beta_{f}$ or in the weak resonance regime, improving DIII-D performance.

\section*{Appendix}

\subsection*{Mode identification}

\begin{figure*}[h!]
\centering
\includegraphics[width=0.8\textwidth]{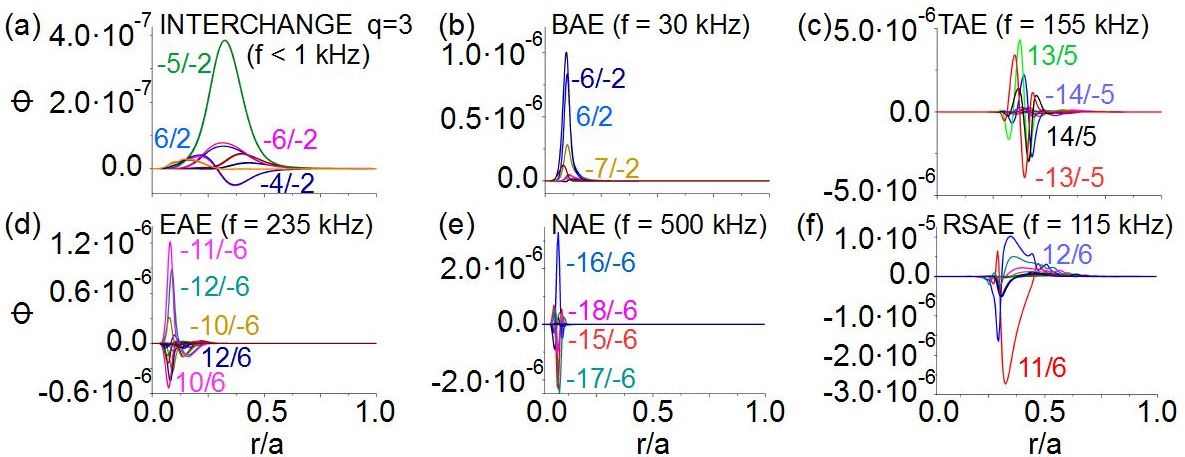}
\caption{$\Phi$ eigenfunctions of a $n=2$ interchange mode (a), $n=2$ BAE (b), $n=5$ TAE (c), $n=6$ EAE (d), $n=6$ NAE (e) and $n=6$ RSAE (f).}\label{FIG:15}
\end{figure*}

Figure~\ref{FIG:15} shows the eigen-function of several instabilities obtained in the simulations. It should be noted that the AEs are primarily identified by the instability position on the Alfv\' en continua (see Figure~\ref{FIG:1}), although some information can also be gained by analyzing the eigenfunction structure, for example from the dominant modes overlapping (overlapped consecutive poloidal modes indicates the destabilization of a TAE) or the eigenfunction peak location along the normalized minor radius (the RSAEs are destabilized near the minimum of the q profile by a single dominant mode). Figure~\ref{FIG:15}a shows a $n=2$ interchange mode (MHD-like instability) destabilized near the q=3 resonant surface during the M1A phase, characterized by a dominant single mode ($-5/-2$) and an instability frequency lower than $1$ kHz. Figure~\ref{FIG:15}b shows a $n=2$ BAE destabilized during the M20 phase, also dominated by a single mode although in this case both parities are important ($6/2$ and $-6/-2$) and the instability frequency is $30$ kHz. Figure~\ref{FIG:15}c shows an $n=5$ TAE unstable in the M10 phase where the $13/5$ and $14/5$ modes are coupled and the instability frequency is $155$ kHz. Figure~\ref{FIG:15}d shows a $n=6$ EAE destabilized during M3C phase where $10/6$ and $12/6$ modes are coupled and the instability frequency is $235$ kHz. Figure~\ref{FIG:15}e shows a $n=6$ NAE destabilized during M2B phase where the modes $-15/-6$ and $-18/-6$ are coupled and the instability frequency is $500$ kHz. Figure~\ref{FIG:15}f shows a $n=6$ RSAE destabilized during M1B phase, where there is a single dominant mode $11/6$ located near the minimum of the reverse shear region with a frequency of $115$ kHz.

\begin{figure*}[h!]
\centering
\includegraphics[width=1.0\textwidth]{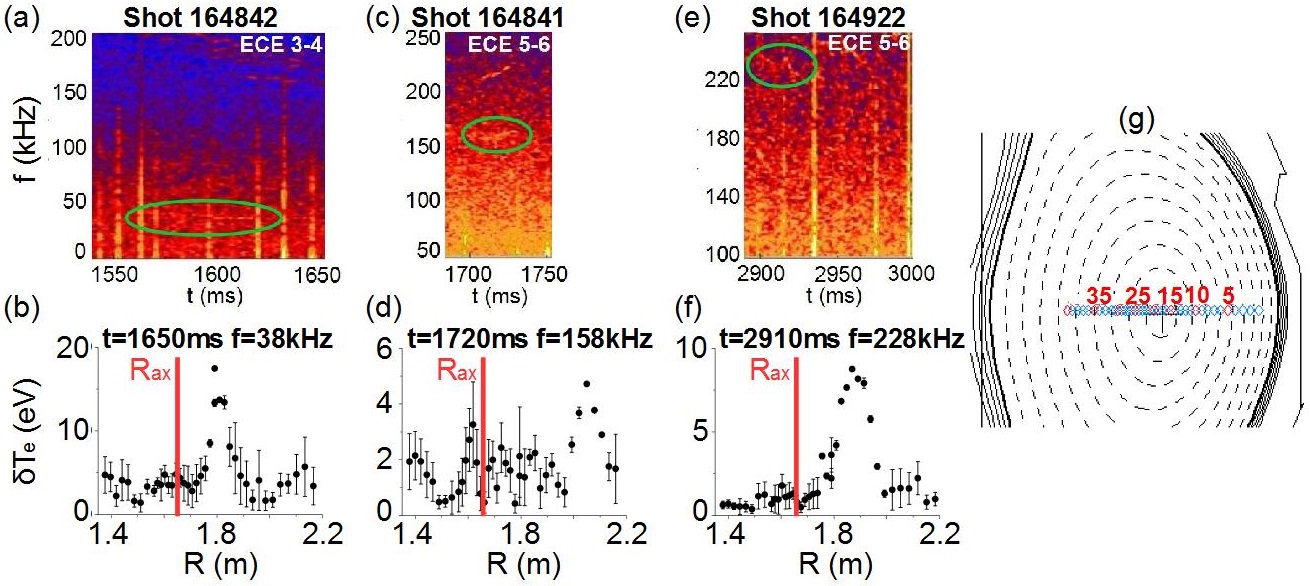}
\caption{ECE data and electron temperature fluctuation of (a and b) shot 164841 at $t=1650$ ms for $f=38$ kHz, (c and d) shot 164842 at $t=1720$ ms for $f=158$ kHz and (e and f) shot 164922 at $t=2910$ ms for $f=228$ kHz. The green oval in the ECE data graphs indicate the instability analyzed. The ECE chords used in the analysis are indicated at the top of the ECE data graphs (white symbols) and the location of the chords is shown in the panel (g). The red solid line in the temperature fluctuation graphs indicates the location of the magnetic axis.}\label{FIG:16}
\end{figure*}

Figure~\ref{FIG:16} shows the Electron Cyclotron emission (ECE) data and the electron temperature fluctuations ($\delta T$). The ECE data is measured along chords that span between $R = [1.4,2.2]$ m. The panels a and b show the data of the shot 164841 at $t=1650$ ms for $f=38$ kHz, indicating the presence of an instability in the range of the $30-40$ kHz located in the inner plasma region, matching the $6/2$ BAE reproduced by the simulations. The panels c and d show the data of the shot 164842 at $t=1720$ ms for $f=158$ kHz, pointing out an instability in the range of the $155-165$ kHz located in the middle-outer plasma region, similar to the $13/5-14/5$ TAE calculated by the simulations although located in the middle plasma region. The panels e and f show the data of the shot 164922 at $t=2910$ ms for $f=228$ kHz, where an instability in the range of the $225-235$ kHz is observed in the inner plasma region, congruent with the $10/6-12/6$ EAE obtained by the simulations. The NAE is not compared with the experimental data because the maximum ECE frequency is limited to $250$ kHz. For the RSAE, the experimental data shows instabilities with the characteristic frequency swiping of the RSAE although at frequencies above $140$ kHz, thus this case is neglected from the comparative study due to that discrepancy.

\ack
This material based on work is supported both by the U.S. Department of Energy, Office of Science, under Contract DE-AC05-00OR22725 with UT-Battelle, LLC and U.S. Department of Energy, Oﬃce of Science, Oﬃce of Fusion Energy Sciences, using the DIII-D National Fusion Facility, a DOE Oﬃce of Science user facility, under Award No. DE-FC02-04ER54698. This research was sponsored in part by the Ministerio of Economia y Competitividad of Spain under project no. ENE2015-68265-P. DIII-D data shown in this paper can be obtained in digital format by following the links at https://fusion.gat.com/global/D3D\_DMP.
\\
\\
DISCLAIMER
\\
This report was prepared as an account of work sponsored by an agency of the United States Government. Neither the United States Government nor any agency thereof, nor any of their employees, makes any warranty, express or implied, or assumes any legal liability or responsibility for the accuracy, completeness, or usefulness of any information, apparatus, product, or process disclosed, or represents that its use would not infringe privately owned rights. Reference herein to any specific commercial product, process, or service by trade name, trademark, manufacturer, or otherwise, does not necessarily constitute or imply its endorsement, recommendation, or favoring by the United States Government or any agency thereof. The views and opinions of authors expressed herein do not necessarily state or reflect those of the United States Government or any agency thereof.

\hfill \break

\end{document}